\documentclass{article}
\usepackage[a4paper, total={6.5in,9in}]{geometry}
\usepackage[utf8]{inputenc}
\usepackage{amsmath}
\usepackage{amssymb}
\usepackage{graphicx}
\usepackage{authblk}
\usepackage{url}
\usepackage[authoryear]{natbib}
\usepackage{algorithm2e}
\usepackage{appendix}
\usepackage{rotating}
\usepackage{appendix}
\usepackage{comment}
\usepackage{longtable}
\usepackage{xcolor}
\usepackage{hyperref}
\usepackage[english]{babel}
\usepackage[autostyle]{csquotes}
\usepackage{setspace}
\usepackage[normalem]{ulem}
\newcommand{\PD}[1]{\textcolor{red}{\textbf{PD}\noindent\textbf{[#1]}}}

\vspace{-30pt}

\author[1]{Pawe{\l} D{\l}otko\thanks{Full Address: Dioscuri Centre in Topological Data Anlaysis, Mathematics Institute, Polish Academy of Sciences, Warsaw, 01-2000, Poland. Email:pdlotko@impan.pl. P.D{\l}otko acknowledges support from the Dioscuri program initiated by the Max Planck Society, jointly managed by the National Science Centre (Poland), and mutually funded by the Polish Ministry of Science and Higher Education and the German Federal Ministry of Education and Research.}}
\affil[1]{Dioscuri Centre in Topological Data Analysis, Polish Academy of Sciences, Poland}
\vspace{-30pt}
\author[2]{Simon Rudkin \thanks{\textbf{Corresponding Author}. Full Address: Institute of Data Science and Artificial Intelligence, School of Social Sciences, University of Manchester, Oxford Road, Manchester, M13 9PL, United Kingdom. Tel: +44 (0)7955 109334 Email:simon.rudkin@manchester.ac.uk}}
\affil[2]{Institute of Data Science and Artificial Intelligence, University of Manchester, UK}
\vspace{-30pt}

\title{Persistence Norms and the Datasaurus}

\begin{document}
\maketitle

\begin{abstract}
    Topological Data Analysis (TDA) provides a toolkit for the study of the shape of high dimensional and complex data. While operating on a space of persistence diagrams is cumbersome, persistence norms provide a simple real value measure of multivariate data which is seeing greater adoption within finance. A growing literature seeks links between persistence norma and teh summary statistics of the data being analysed. This short note targets the demonstration of differences in the persistence norms of the Datasaurus datasets of \cite{matejka2017same}. We show that persistence norms can be used as additional measures that often discriminate datasets with the same collection of summary statistics. Treating each of the data sets as a point cloud we construct the $L_1$ and $L_2$ persistence norms in dimensions 0 and 1. We show multivariate distributions with identical covariance and correlation matrices can have considerably different persistence norms. Through the example, we remind users of persistence norms of the importance of checking the distribution of the point clouds from which the norms are constructed.
\end{abstract}
\section{Introduction}

Topological Data Analysis (TDA) concerns the study of the joint distribution of ordinal data. The TDA toolkit is equipped with many measures, such as the persistence norm, which allow the user to gain a summary of data shape. The datasaurus dozen of \cite{matejka2017same} is presented as an important reminder that the mean, correlation and standard deviation of a bi-variate dataset do not provide sufficient information to understand the appearance of the scatter plot. In the language of TDA, the scatter plot of two variables may be thought of as a point cloud in which co-ordinates on the two axes define the locations of each point within the cloud. Through the calculation of the $L_1$ and $L_2$ persistence norms in dimensions 0 and 1 we demonstrate that the persistence norms do offer additional information about the point clouds. We show norms contain information over and above the means, variances and pairwise correlations of the constituent variables. Whilst the datasaurus dozen datasets satisfy the ``same stats different graphs'' requirement \citep{matejka2017same}, they have considerably different persistent norms. 

Further value in the consideration of the datasaurus dozen lies in the ability to distinguish between datasets which offer higher $L_1$ and $L_2$ norms in dimension 1. $L_1$ and $L_2$ norms are often highly correlated in applications of TDA, rendering the consideration of cases where the two norms differ of additional importance. Where there are differentials in the two norms, further information is revealed from the dataset. $L_1$ norms are shown to reach a maximum when there are many smaller features in the dataset, whilst the highest $L_2$ norms occur when there are larger single features created by the points. Consequently, the datasaurus example acts to remind that although $L_1$ and $L_2$ norms are highly correlated, there is inference to be taken from any difference in their ratio. Where the choice is made to consider either $L_1$ or $L_2$ norms in empirical applications of TDA this note is a reminder that the decision is not made free of charge. 

A final contribution from the calculation of the persistence norms of the datasaurus datasets is evidence that just looking at the $L_1$ and $L_2$ norms as additional measures on top of the standard summary statistics is still not sufficient to fully classify the different data sets. We show that many of the datasets have very similar norms. By adding the maximum distance between a pair of points in the dataset there is sufficient information to classify the data sets\footnote{The maximal distance between two points within the cloud is a very unstable measure. A single outlier point within the data can kill the validity of the measure.}. However, with only 14 data sets considered in this paper it is only an exercise to demonstrate classification. In the practical implementation of persistence norms, far greater value is derived from the additional information embedded within the norms relative to correlations and variances, than is gained from metrics such as maximal separation between points. As the adoption of persistence norms accelerates, so it is important to understand how these norms relate to data shape.

Extending the discussion beyond the basic datasaurus datasets, we show the impact of a constant scaling being applied to all variables is simply to scale the persistence norms by the same constant.  Translations, rotations or any other affine transformations of the point cloud do not alter the shape and therefore do not alter the persistence norms. When the cloud is scaled by different amounts on each axis the resultant distortion means the effect on the persistence norm is non-linear and different for each dataset. By plotting univariate distributions we learn more about the differences between the considered datasets, underlining the value in the metrics on the shape of multi-dimensional data provided by TDA. In all each extension the notion of the same first and second order moments is not violated, all differences in persistence norms continue to be understood through the differences in the shape of the point cloud.

This note contributes to a growing literature which seeks to understand how persistence norms relate to understood financial metrics. Following the early demonstration that persistence norms can act as an early warning signal for financial crashes \citep{gidea2018topological,gidea2020topological} there has been a quest to understand why. \cite{aromi2021topological} considers the role of the covariance matrix in determining norms and \cite{katz2022topological} looks at the impact of noise on persistence norms. \cite{rudkin2023topology} evidences that correlation and volatility alone do not explain persistence norms in cryptocurrency markets. \cite{rudkin2023uncertainty} shows that there are links between persistence norms and uncertainty, but that there is still an unexplained component in the norms of financial markets. \cite{akingbade2023topological} updates the discussion of \cite{gidea2018topological}, linking crashes to periodicity\footnote{See \cite{dlotko2019cyclicality} for a demonstration of the ability of TDA to detect periodic behaviour earlier than other widely applied methodologies.}. The question of how datasets with identical correlations and volatilities (standard deviations) can have different norms remains open.  As we demonstrate in this note, distribution is a very important component of the link between norms and the summary statistics of data.

To accompany this note we provide a full R code which can be accessed through the website of Simon Rudkin\footnote{Access is via either \href{https://sites.google.com/view/simonrudkin/home}{https://sites.google.com/view/simonrudkin/home}}. If using the code then please do not forget to cite this note and the R packages used therein.

\section{Persistence Norms}
\label{sec:norms}

The process of construction of persistence norms typically begins with a point cloud. In two dimensions the point cloud is simply the scatter plot. The co-ordinates of a point within the cloud are determined by the values of that point on each of the considered axes. The Datasaurus dozen are of interest because their respective point clouds all differ despite the fact that the means and standard deviations of the two variables that provide the axes, and correlations between the two axis variables, are identical. Herein we give an illustrative example for the construction of persistence norms. A fuller description of the mathematics underlying the diagrammatic illustration may be found in \cite{rudkin2023topology}.

\begin{figure}

	\begin{center}
		\caption{Construction of Dimension 1 Homology Classes}
		\label{fig:calcnorm}
		\begin{tabular}{c c c}
			\includegraphics[width=5cm]{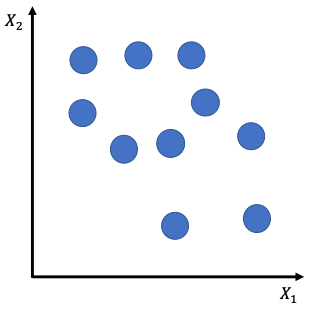}&
			\includegraphics[width=5cm]{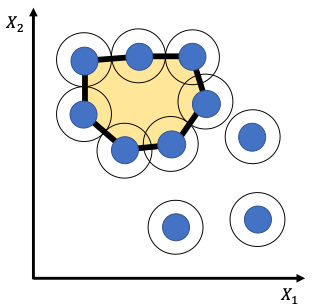}&
			\includegraphics[width=5cm]{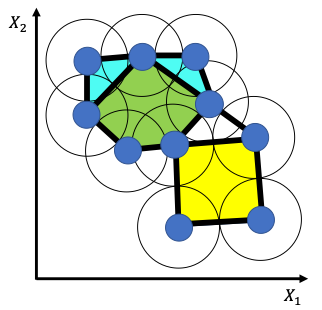}\\
			(a) Point cloud & (b) Filtration 1 ($\epsilon_1$) & (c) Filtration 2 ($\epsilon_2$)
		\end{tabular}
	\end{center}
	\raggedright
	\footnotesize{Notes: Figures demonstrate the construction of persistence norms for an artificial data set with two variables, $X_1$ and $X_2$. Panel (a) is a scatter plot of the point cloud. A filtration of $\epsilon_1$ is applied in panel (b). To see the filtration circles of radius $\epsilon_1$ are drawn centred on each of the data points. Where the circles touch, or there is overlap, an edge is constructed between the points. A combination of edges is a dimension 0 feature. Where there exists an area enclosed by four or more points a shape is formed. In panel (a) the single shape is coloured orange. A shape formed within at least four points is a feature in dimension 1. Panel (c) applies a filtration of $\epsilon_2$. An edge is a dimension 0 feature. Where two dimension 0 features meet the longest lived feature remains and the shortest lived feature dies. A smaller shape is formed coloured green. Another shape is also born in the lower right coloured yellow.}
\end{figure}

Figure \ref{fig:calcnorm} demonstrates the process through which dimension 1 homology classes are formed. Through considering the dimension 1 homology classes the construction of the dimension 1 persistence norms is understood. In panel (a) the dataset is shown as a scatterplot with $X_1$ on the horizontal axis and $X_2$ on the vertical. In panel (b) a filtration of $\epsilon_1$ is applied and this is shown as a larger circle around each data point. Where those circles touch an edge is formed between the two data points. Edges are shown as thicker black lines. Where four or more points sit on the perimeter of a shape which does not have any edges crossing its interior, the resulting shape is a feature in dimension 1. In panel (b) there is a single dimension 1 feature. The birth of this feature is $\epsilon_1$ as there are many edges on the outside of the shape which are formed by the circles just touching. In panel (c) a larger filtration, $\epsilon_2$, is applied. Two edges now form across the corners of the original orange shape. These edges mean that the orange shape is no longer a dimension 1 feature and so that feature dies. New features are formed by the creation of the smaller green shape and a further feature in the lower right of the shape. These green and yellow coloured features in panel (c) thus have a birth filtration of $\epsilon_2$. 

Dimension 0 features are based upon the connected components. Hence, they are merged by edges which form between the points. Firstly, zero dimensional features are supported in the data points themselves, this exists with a filtration 0. Subsequently, with an increasing filtration points are merged until only a single connected component is left. The obtained single connected component will always be there and so the first dimension 0 feature is infinitely lived. Whenever an edge forms a dimension 0 feature dies. Where two features join by an edge then we consider that the one which has the highest birth filtration dies and the feature with the earlier birth filtration, the longest-lived feature, continues. In Figure \ref{fig:calcnorm}, Panel (a) we see 10 different connected components. Panel (b) shows four connected components, after seven initial ones are merged into one. In Panel (c) the final signle 0-dimensional feature is presented. Closer inspection will reveal that there are many balls which just touch in panel (b). For a filtration slightly below $\epsilon_1$ there will have been more dimension 0 features.

Given persistence diagrams in different dimensions, the norms for dimension $g$, $g \in \lbrace 0,1 \rbrace$ are computed using: 

\begin{align}
    L_{g1} = \sum_{f=1}^{F_g} d_f-b_f \\ \label{eq:l1}
\end{align}
and:
\begin{align}
    L_{g2} = \sqrt{\sum_{f=1}^{F_g} (d_f-b_f)^2 } \label{eq:l2}
\end{align}
Where in both cases, the second digit in the subscript informs that we are calculating the $L_1$ or $L_2$ norm. $F_g$ here denotes the number of features in dimension $g$. The lifetime for each feature, $f$, in dimension $g$ is computed using $d_f$ and $b_f$, being the birth, and death filtrations of feature $f$. 

Within this note the persistence norms are constructed using the TDA package in R \citep{TDAR2022}. Full details of the technical implementation may be found in the documentation accompanying the package.

\section{Data}

Data used in this note is from \cite{matejka2017same} and is included within the accompanying R package \citep{datasauRus2022}. Each dataset includes points described by two variables, $X_1$ and $X_2$ and is designed to have identical first and second moments. That is the average values of $X_1$ and $X_2$, $\mu_1$ and $\mu_2$ respectively, are identical. Likewise the standard deviations of both variables, $\sigma_1$ and $\sigma_2$ are also identical. The correlation between $X_1$ and $X_2$, $\rho_{12}$ is equal for every dataset. Specifically the datasets are created such that $\mu_1 = 54.27$, $\mu_2=47.84$, $\sigma_1=16.77$, $\sigma_2 =26.94$ and $\rho_{12}=-0.064$. Table \ref{tab:sumstat} provides summary statistics for the datasets to confirm the equality across all sets. For simplicity of exposition we refer to these datasets as having the same summary statistics.

\begin{table}
	\begin{center}
		\caption{Datasaurus Data Summary Statistics}
		\label{tab:sumstat}
		\begin{tabular}{l l c c c c c c  c}
			\hline
			Dataset & Var& Mean & s.d. & Min & q25 & q50 & q75 & Max\\
			\hline
			Dino & $X_1$&54.26&16.77&22.31&44.10&53.33&64.74&98.21\\
			&$X_2$&47.83&26.94&2.95&25.29&46.03&68.53&99.49\\
			Normal &$X_1$&54.26&16.77&9.18&42.42&57.43&66.91&100.05\\
            & $X_2$ &47.83&26.93&3.64&24.45&46.54&67.97&106.38\\
			Away & $X_1$ &54.27&16.77&15.56&39.72&53.34&69.15&91.64\\
			&$X_2$&47.83&26.94&0.02&24.63&47.54&71.80&97.48\\
			Bullseye & $X_1$&54.27&16.77&19.29&41.63&53.84&64.80&91.74\\
			&$X_2$&47.83&26.94&9.69&26.24&47.38&72.53&85.88\\
			Circle &$X_1$ &54.27&16.76&21.86&43.38&54.02&64.97&85.66\\
			&$X_2$&47.84&26.93&16.33&18.35&51.03&77.78&85.58\\
			Dots &$X_1$&54.26&16.77&25.44&50.36&50.98&75.20&77.95\\
			&$X_2$&47.84&26.93&15.77&17.11&51.30&82.88&94.25\\
			H Lines & $X_1$&54.26&16.77&22.00&42.29&53.07&66.77&98.29\\
			&$X_2$&47.83&26.94&10.46&30.48&50.47&70.35&90.46\\
			High Lines & $X_1$&54.27&16.77&17.89&41.54&54.17&63.95&96.08\\
			&$X_2$&47.84&26.94&14.91&22.92&32.50&75.94&87.15\\
			Slant Down &$X_1$&54.27&16.77&18.11&42.89&53.14&64.47&95.59\\
			&$X_2$&47.84&26.94&0.30&27.84&46.40&68.44&99.64\\
			Slant Up &$X_1$&54.27&16.77&20.21&42.81&54.26&64.49&95.26\\
		    &$X_2$&47.83&26.94&5.65&24.76&45.29&70.86&99.58\\
			Star &$X_1$&54.27&16.77&27.02&41.03&56.53&68.71&86.44\\
			&$X_2$&47.84&26.93&14.37&20.37&50.11&63.55&92.21\\
			V Lines & $X_1$ &54.27&16.77&30.45&49.96&50.36&69.50&89.50\\
			&$X_2$&47.84&26.94&2.73&22.75&47.11&65.85&99.69\\
			Wide Lines &$X_1$&54.27&16.77&27.44&35.52&64.55&67.45&77.92\\
			&$X_2$&47.83&26.94&0.22&24.35&46.28&67.57&99.28\\
			X Shape &$X_1$&54.26&16.77&31.11&40.09&47.14&71.86&85.45\\
			&$X_2$&47.84&26.93&4.58&23.47&39.88&73.61&97.84\\
			
			\hline
		\end{tabular}
	\end{center}
\raggedright
\footnotesize{Notes: Figures report the means, standard deviations, minimum, maximum and quartiles of the variables included within the Datasaurus datasets of \cite{matejka2017same}. Dino is the original datasaurus, Normal is an additional dataset created for this paper which is comprised of two random normal variables with the same summary statistics and correlation as the datasaurus. The remaining 12 datasets are the datasaurus dozen of \cite{matejka2017same}.}
\end{table}

To those included in \cite{matejka2017same} we add a further dataset in which both $X_1$ and $X_2$ are random draws from normal distributions. The Normal dataset has $X_1 \sim N(\mu_1,\sigma_1^2)$ and $X_2 \sim N(\mu_2,\sigma_2^2)$. As with the datasaurus dozen datasets, correlation between $X_1$ and $X_2$ is also -0.064. In many statistical applications we make the assumption of normally distributed variables and hence the Normal dataset is a useful example to present alongside the datasaurus dozen \citep{matejka2017same}. In total we have 14 datasets 

\begin{figure}
	\begin{center}
		\caption{Scatter Plots}
		\label{fig:scatter}
		\begin{tabular}{c c c c}
			\multicolumn{2}{c}{\includegraphics[width=7cm]{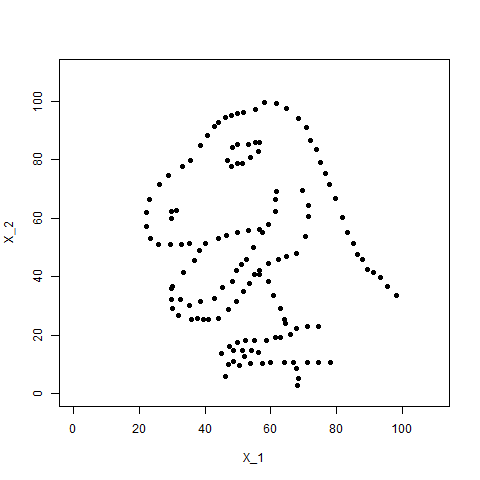}}&
			\multicolumn{2}{c}{\includegraphics[width=7cm]{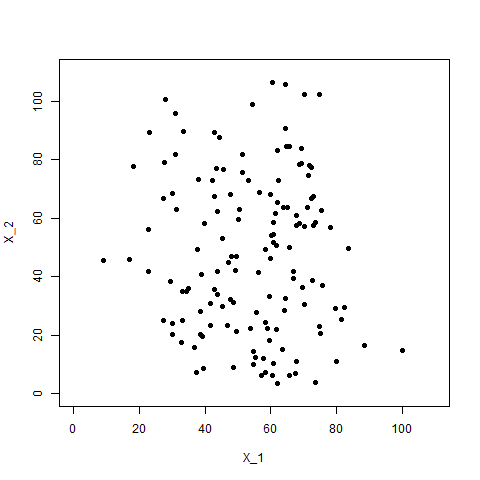}}\\
			\multicolumn{2}{c}{(a) Dino} &
			\multicolumn{2}{c}{(b) Normal} \\
			\includegraphics[width=3.5cm]{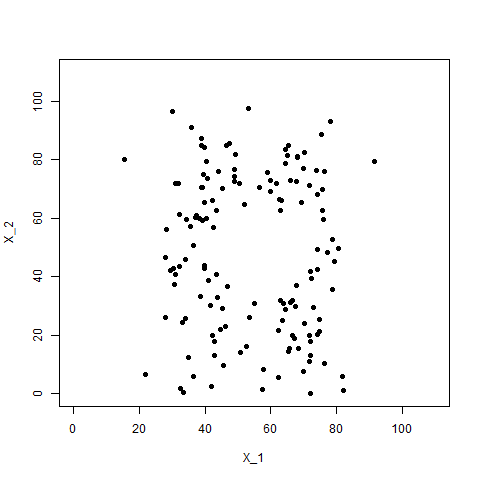}&
			\includegraphics[width=3.5cm]{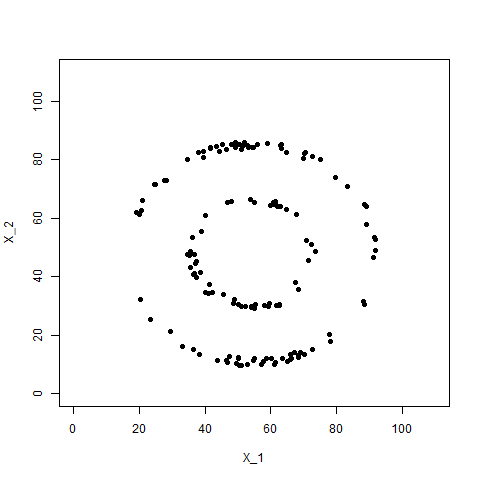}&
			\includegraphics[width=3.5cm]{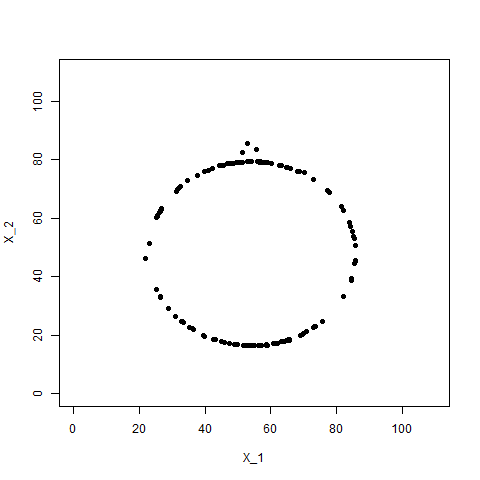}&
			\includegraphics[width=3.5cm]{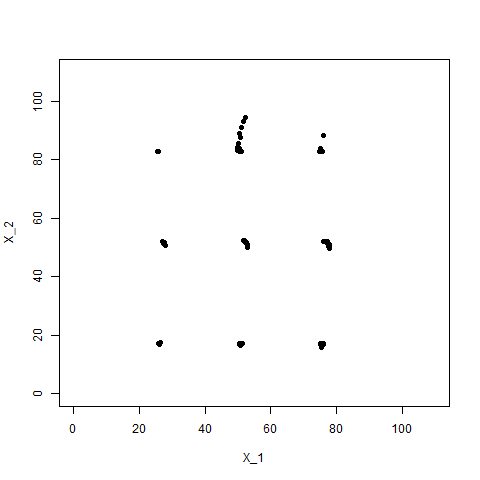}\\
			(c) Away & (d) Bullseye & (e) Circle & (f) Dots \\
			\includegraphics[width=3.5cm]{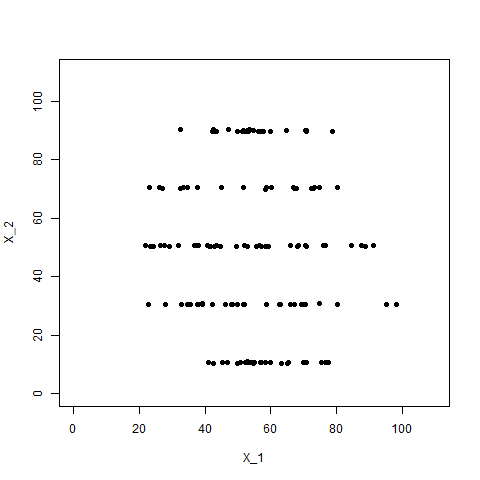}&
			\includegraphics[width=3.5cm]{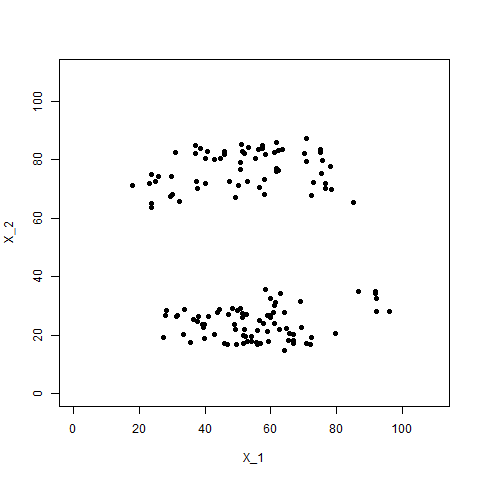}&
			\includegraphics[width=3.5cm]{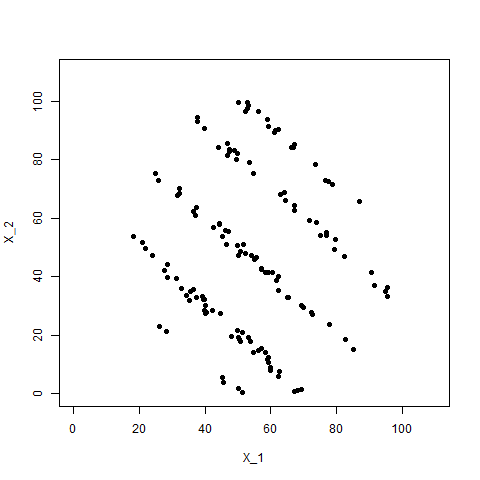}&
			\includegraphics[width=3.5cm]{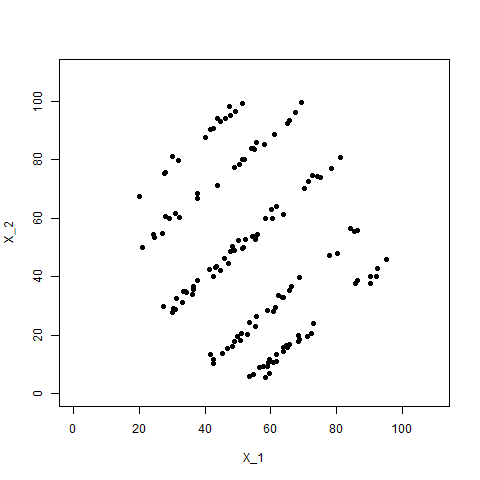}\\
			(g) H lines & (h) High lines & (i) Slant down & (j) Slant up \\
			\includegraphics[width=3.5cm]{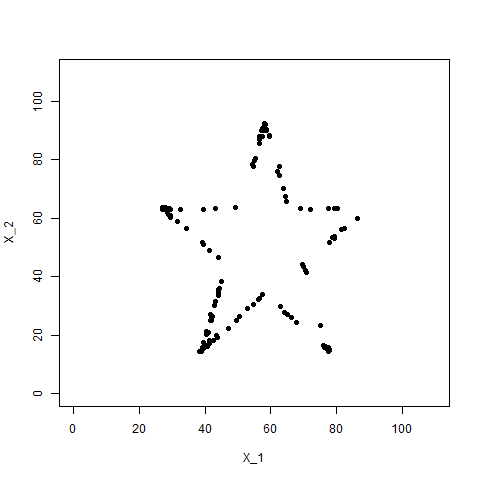}&
			\includegraphics[width=3.5cm]{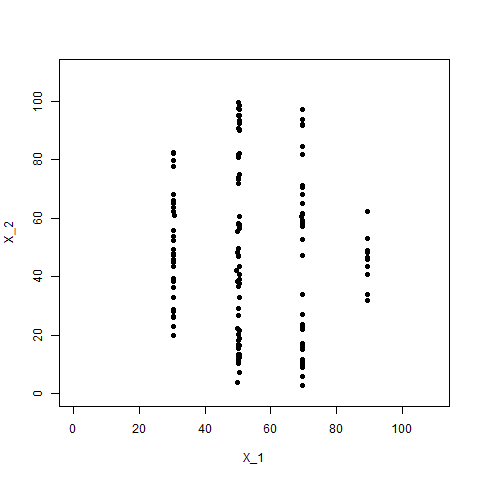}&
			\includegraphics[width=3.5cm]{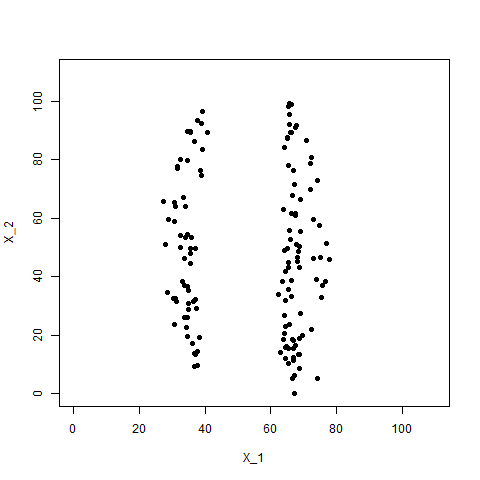}&
			\includegraphics[width=3.5cm]{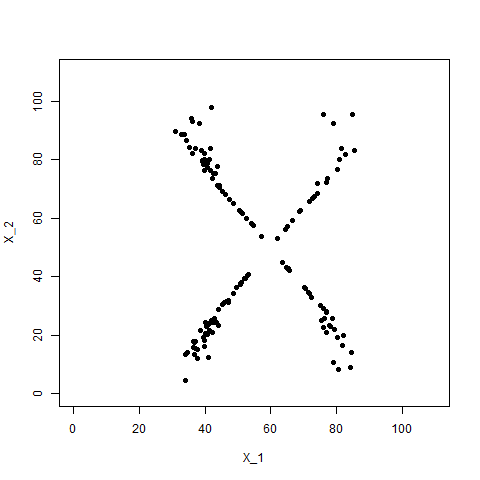}\\
			(k) Star & (l) V lines & (m) Wide lines & (n) X shape \\
		\end{tabular}
	\end{center}
\footnotesize{Notes: Scatterplots of the original datasaurus dataset, the Normal cloud constructed in this paper and the datasaurus dozen of \cite{matejka2017same}. All clouds are constructed from two variables, $X_1$ and $X_2$ with identical means, $\mu_1$, $\mu_2$, and standard deviations, $\sigma_1$, $\sigma_2$. For all datasets $\mu_1 = 54.27$, $\mu_2 = 47.84$, $\sigma_1 = 16.77$ and $\sigma_2 = 26.94$. Correlation between the two axis variables, $\rho_{12}$, is also identical in every panel. In all cases $\rho_{12} = -0.064$.}
\end{figure}

Figure \ref{fig:scatter} shows the original datasaurus in panel (a). The newly constructed normal dataset is shown in panel (b). The remaining 12 panels are the datasaurus dozen as constructed in \cite{matejka2017same}. There are groups of pictures, for example the H lines in panel (g) and the V lines in panel (l). Panels (i) and (j) present a direct comparison between the Slant down and the Slant up. There are large contrasts between the evident shapes in the Bullseye of panel (d), the Star of panel (k) and the X shape of panel (n). The datasaurus itself, Dino, was designed as a picture to show the importance of seeing the shape of data. 

Formally we may consider many of the datasets within the \cite{matejka2017same} set as being approximate rotations of each other. High lines in panel (h) of Figure \ref{fig:scatter} and Wide lines in panel (m) are close to being a 90 degree rotation of each other. Likewise, H lines in panel (g) is a rotation of V lines in panel (l), and slant down in panel (i) is a rotation of Slant up in panel (j). All of these six panels may be thought of as variations of the arrangement of points in lines. Panels (h) and (m) have just two lines of points. Panels (i) and (j) both have 5 lines. Because of the different mean and standard deviation for $X_1$ and $X_2$, panel (g) and (l) have the greatest contrast of the rotation pairs.

Panels (c), (d) and (e) in Figure \ref{fig:scatter} are all variations on having a large hole in the middle of the data. In panel (c), the Away dataset, the hole is surrounded by a noisy set of points which spread out through the space. The Bullseye dataset in panel (d) has gaps in the rings, but all points are very close to the two elliptical shapes that form the ''eye''. Finally in the Circle dataset there is just one ring of tightly arranged points around a larger central hole\footnote{The use of the word circle to describe this data set fits from the fact that the scatter plot does have a circular shape. Thinking of the hole as an ellipse does not change the inference.}. There are three points forming a shape like the stone on a ring at the top of the circle. These three points are the only divergence from the perfect circle.

Cases where the plots are variants of each other further reinforce the need to view the picture to distinguish between the cases. Albeit with slight movement to maintain the overall correlation, the rotated datasets have similar distances between points. Consequently we would expect similar persistence norms for these rotated datasets. Given the stated aim is not to classify the datasets, the similarity of panels provides an excellent opportunity to demonstrate consistency of persistent norms independent of the axes. In the later analysis, a maximal distance between points is included to underline the similarity between rotated datasets. The maximal distance adds confirmation of similarity between rotated datasets alongside that which comed from the rotated datasets hvaing similar norms. 

\section{Results}
\label{sec:results}

\subsection{Persistence Diagrams}

Calculation of the persistence norms is based upon the identification of features in dimensions 0 and 1. Each feature is created at a certain value of filtration called the birth value of the feature and cease to exist at a death value of filtration. These birth and death filtrations may be plotted upon a persistence diagram. A persistence diagram plots the birth on the horizontal axis and the death on the vertical. All points must lie above the 45 degree line since the birth filtration must be lower than the death. We may also estimate a boundary above the 45 degree line based on the probability of the identified feature being non-random. The bootstrap technique to identify the confidence interval follows \cite{fasy2014confidence}. We do not discount any features in the confidence interval from the persistence norm calculations, but it is helpful to understanding the dataset to see how many features exist outwith the confidence interval.

\begin{figure}
    	\begin{center}
		\caption{Persistence Diagrams}
		\label{fig:pdiag}
		\begin{tabular}{c c c c}
			\multicolumn{2}{c}{\includegraphics[width=7cm]{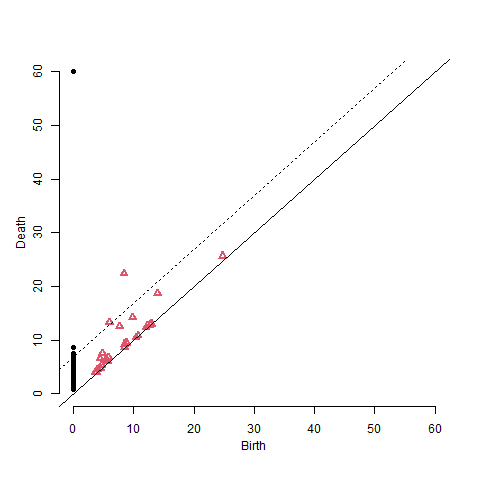}}&
			\multicolumn{2}{c}{\includegraphics[width=7cm]{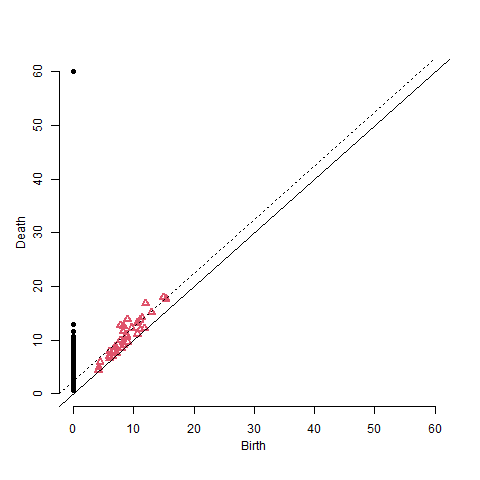}}\\
			\multicolumn{2}{c}{(a) Dino} &
			\multicolumn{2}{c}{(b) Normal} \\
			\includegraphics[width=3.5cm]{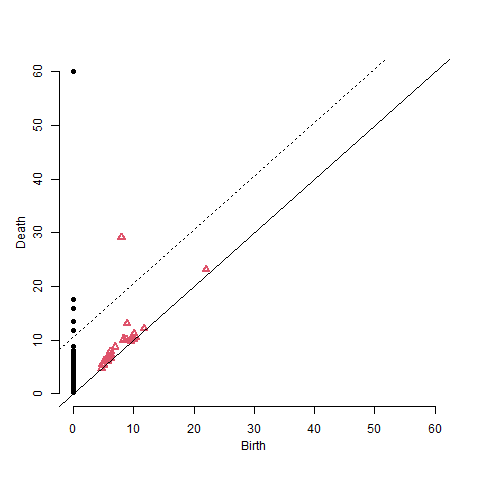}&
			\includegraphics[width=3.5cm]{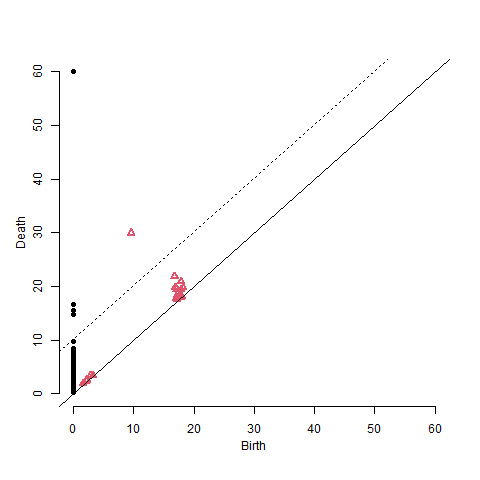}&
			\includegraphics[width=3.5cm]{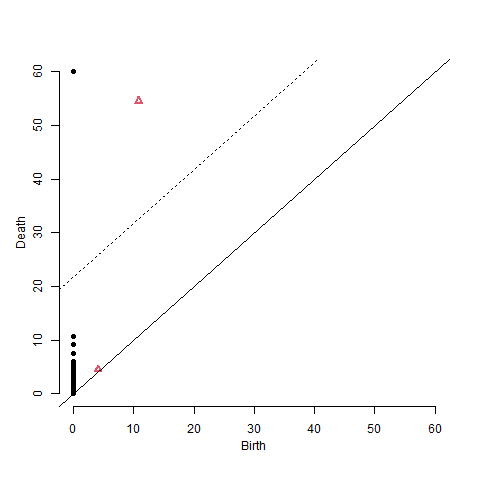}&
			\includegraphics[width=3.5cm]{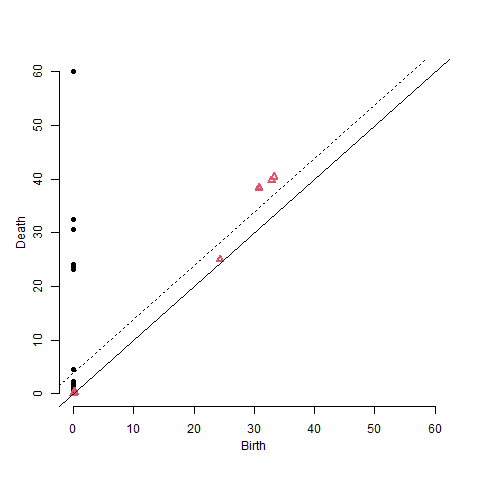}\\
			(c) Away & (d) Bullseye & (e) Circle & (f) Dots \\
			\includegraphics[width=3.5cm]{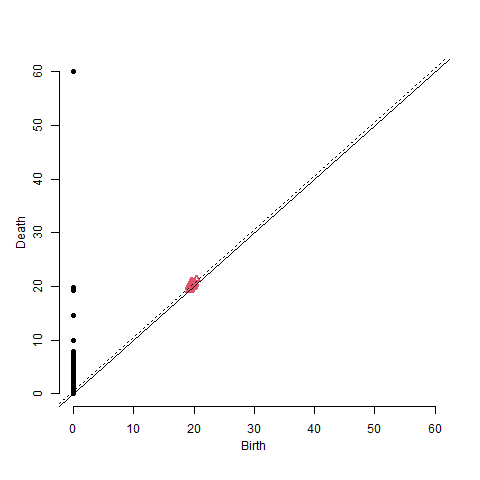}&
			\includegraphics[width=3.5cm]{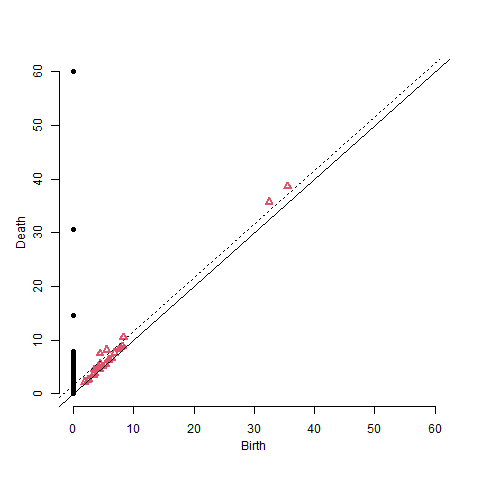}&
			\includegraphics[width=3.5cm]{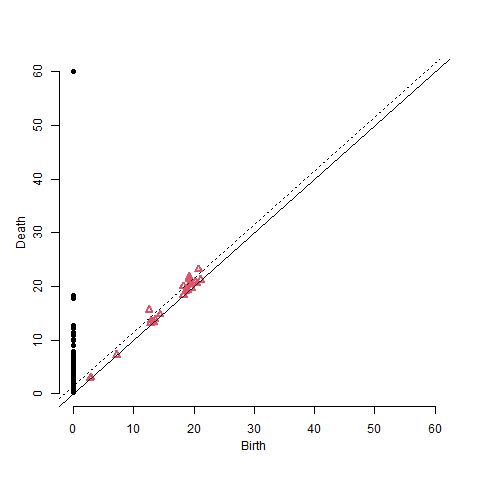}&
			\includegraphics[width=3.5cm]{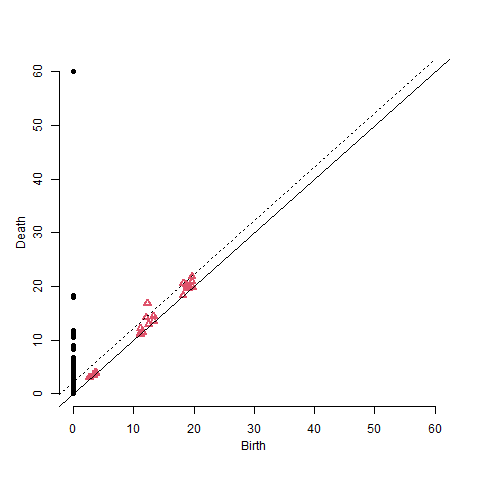}\\
			(g) H lines & (h) High lines & (i) Slant down & (j) Slant up \\
			\includegraphics[width=3.5cm]{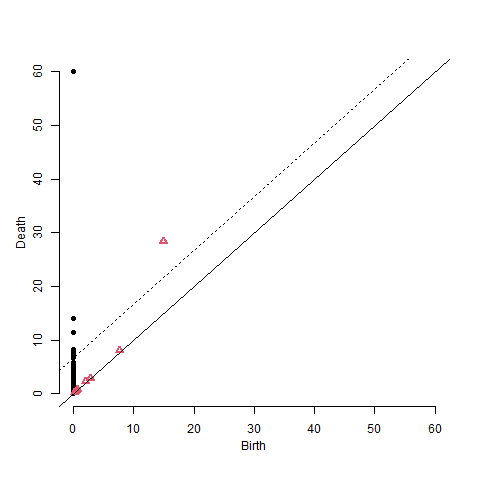}&
			\includegraphics[width=3.5cm]{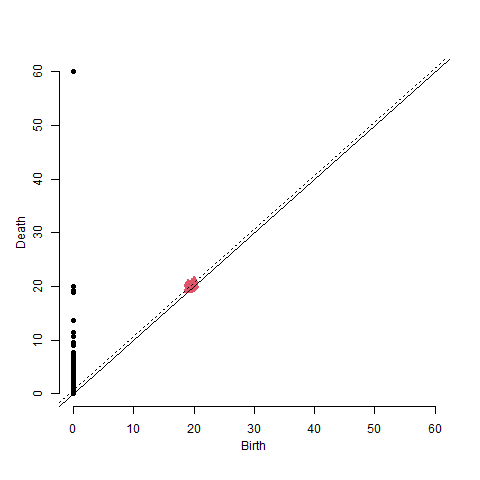}&
			\includegraphics[width=3.5cm]{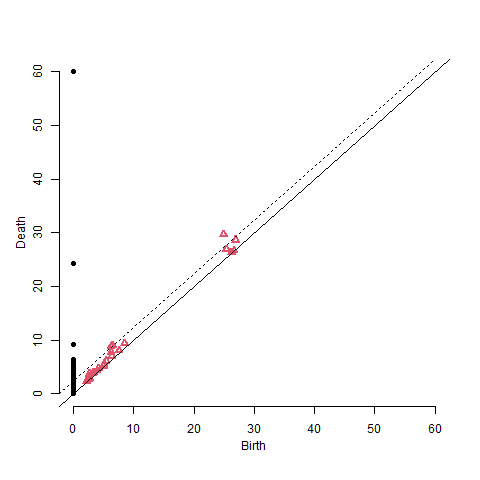}&
			\includegraphics[width=3.5cm]{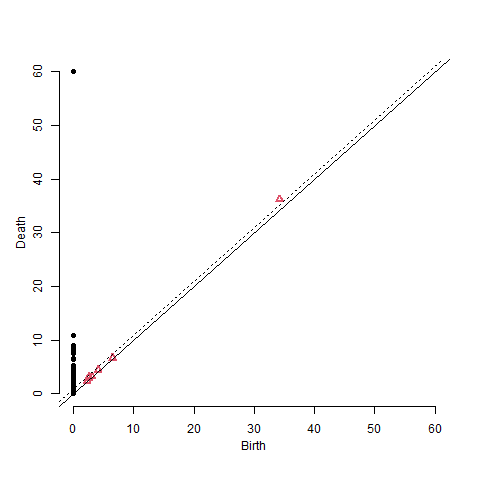}\\
			(k) Star & (l) V lines & (m) Wide lines & (n) X shape \\
		\end{tabular}
	\end{center}
\footnotesize{Notes: Persistence diagrams for the original datasaurus dataset, the normal cloud constructed in this paper and the datasaurus dozen of \cite{matejka2017same}. All clouds are constructed from two variables, $X_1$ and $X_2$ with identical means, $\mu_1$, $\mu_2$, and standard deviations, $\sigma_1$, $\sigma_2$. For all datasets $\mu_1 = 54.27$, $\mu_2 = 47.84$, $\sigma_1 = 16.77$ and $\sigma_2 = 26.94$. Correlation between the two axis variables, $\rho_{12}$, is also identical in every panel. In all cases $\rho_{12} = -0.064$.}
\end{figure}

Figure \ref{fig:pdiag} contains the 14 persistence diagrams for the datasets considered in this note. The original datasaurus and the normal dataset are placed in panels (a) and (b) respectively. The ``datasaurus dozen'' of \cite{matejka2017same} then correspond to panels (c) to (n). Immediately the contrast between the Dino and Normal datasets may be seen. The Dino has more dimension 1 features which sit outside the confidence interval. For the Normal plot there are more dimension 1 features, but most are within the confidence interval. The dimension 0 features, coloured black on the persistence diagram, show a bigger range of death filtrations for the normal distribution than do those in the dinosaur. Inspection of the dataset in panels (a) and (b) of Figure \ref{fig:scatter} reveals that points in the dinosaur are closer to each other and hence dimension 1 features connect with others at similar filtration radii. Consequently, the persistence diagrams inform that differentials in the shape of data will appear in the calculated persistence norms.

Panels (c) to (n) show three patterns. Firstly, there are those datasets where there is a dimension 1 feature which has a long life, being born at low filtration and then dying at a high filtration. Panels (c), (d), (e) and (k) have this feature. The long-lived dimension 1 feature in panel (e) is particularly long-lived. In practice, panel (e) is the easiest to understand because the hole is the middle of the circle of points. The centre of the circle only disappears when the filtration is half the radius. The large holes in the centres of panels (c) and (d) may also be seen when looking at the scatter plots in Figure \ref{fig:scatter}. For panel (k), the Star, the hole is the area enclosed within the Star, but the closeness of the points in the centre means that the hole is effectively much smaller than the distance between the points of the Star. Indeed the two sides of each point of the Star will connect at low filtrations and it is only when the central hole forms that the long-lived feature is born. In panel (k) of Figure \ref{fig:pdiag} we see that the longest lived feature is born at a higher filtration than (c), (d) or (e). Amongst these four panels with a long lived dimension 1 features, the Circle and Star diagrams have unique properties. For the Away and the Bullseye data sets, the former has more short-lived features at lower radii, whilst the latter has a set of features which form at higher radii. In practice the difference is that the Away dataset has more spread between points in general, whilst the Bullseye has points which are more carefully arranged and hence, central hole aside, only form shapes when there are connections between the inner and outer ring. 

The second notable pattern in the persistence diagrams occurs where there are short-lived features which form at high filtrations. Examples of this pattern may be found in panels (f), (g), (h), (l), (m) and (n). These are patterns where there are large gaps between points which become connected to create the dimension 1 feature. In panel (f), Dots, the rationale is clearest. Panel (f) shows 9 small clusters of points which are arranged in a square pattern. As the edges of the square form we see 4 dimension 1 features emerge. Only when the diagonals cut across each of those small squares will the dimension 1 feature die. Consequently there are 4 short lived features forming at high filtration. In panels (f) to (h), as well as (l) to (n), there are large gaps between lines of points. If the lines are perfectly evenly spaced then all connections across the gaps will form at the same filtration, then the features will die when a diagonal connection forms across them. We do see some evidence of this late birth then subsequent early death driving the very short lived features. There are also gaps between points in the lines seen in most of the panels. These gaps between points in the line may also create shorter lived features in the same way. If the first connections create a shape then that feature lives until a further connection closes the gap. These closures may happen very quickly, depending on the spacing of the points either side of the gap. Where features have very short lives, the influcence on the nortms is small. $L_2$ norms are comparatively unaffected since they are based on the sum of squared lives and the closures of gaps create features with very short lives. 

The third pattern is displayed in panels (i) and (j) is two groups of short lived features. These panels represent the datasets Slant up and Slant down, and so understandably share many similarities. Unlike the H lines and V lines datasets, the Slant up and Slant down datasets do not have evenly spaced lines of points. The result of the spacing being uneven is that there are dimension 1 features formed from connections between the closer pairs of parallel lines of points first, and then dimension 1 features formed from the further pairs of parallel lines of points second. Hence we see clusters of features on the persistence diagram at a similar filtration to half the distance between the closer pairs of lines, and then a second cluster at a similar filtration to half the distance between the further lines. Recall that an edge forms between two points when the two ``balls'' surrounding that pair of points touch. Hence the distance between the points would be twice the radius of the ball; the ball radius is half the distance between the points. At this point we note that there are strong similarities between the persistence diagrams in panels (i) and (j) of Figure \ref{fig:pdiag}. Any differential is down to the precise location of the points and the need to maintain the loose negative overall correlation between the points within a dataset. 

Across the three patterns outlined, we may understand the rationales for the observed differences in the dimension 1 persistence norms. The persistence diagrams show dimension 0 features as black dots. Because each data point is a feature, we see that all have a birth filtration on 0. Where a connection forms between two points that connection assumes the birth of the longest lived feature, that is 0. Close inspection of the plots shows different patterns within the black points Dots in panel (f) is notably spread compared to the Dino in panel (a). Higher deaths are also observed for some features in the Normal plot, panel (b), versus the Star in panel (k). Because the dimension 1 features are easier to observe in both the persistence diagrams and scatter plots, we do not present more detail on the dimension 0 differentials here. We may also understand that there differences in the dimension 0 persistence norms which may be used to understand the datasets. 

\subsection{Persistence Norms}

The persistence diagrams in Figure \ref{fig:pdiag} show the births and deaths of the various features in dimensions 0 and 1. From the births and deaths we compute the persistence norms using equations \eqref{eq:l1} and \eqref{eq:l2}. We also capture the spread of the data points through the minimum and maximum values on each axis and then take the maximum distance between data points. Table \ref{tab:norms} also includes the standard deviations and correlations for ease of comparison with the additional metrics.

\begin{sidewaystable}
	\begin{center}
		\caption{Datasaurus Datasets Summary Statistics with Persistence Norms}
		\label{tab:norms}
		\begin{tabular}{l l l c c c c c c c c c c c}
			\hline
			Dataset & Panel & $L_{01}$&$L_{02}$&$L_{11}$&$L_{12}$& $\sigma_1$ & $\sigma_2$ & $\rho_{12}$& $Min(X_1)$& $Max(X_1)$& $Min(X_2)$& $Max(X_2)$ & $Dist$\\
			\hline
			Dino& (a) & 485.707&43.615&47.334&18.157&16.765&26.935&-0.064&22.308&98.205&2.949&99.487&97.082\\
			Normal&(b)&644.139&61.632&63.98&13.548&16.77&26.93&-0.064&9.185&100.049&3.635&106.379&112.12\\\\
			Away & (c) &581.987&58.428&44.079&22.067&16.77&26.94&-0.064&15.561&91.64&0.015&97.476&108.686\\
			Bullseye & (d) &379.038&45.795&44.096&21.865&16.769&26.936&-0.069&19.288&91.736&9.692&85.876&76.824\\
			Circle & (e)&198.978&27.085&44.136&43.76&16.76&26.93&-0.068&21.864&85.665&16.327&85.578&69.342\\
			
			Dots&(f)&241.529&73.71&29.899&14.633&16.768&26.93&-0.06&25.444&77.954&15.772&94.249&87.128\\
			H lines &(g)&366.722&53.457&10.532&2.558&16.766&26.94&-0.062&22.004&98.288&10.464&90.459&91.601\\
			High lines &(h)&423.415&50.484&23.147&6.921&16.767&26.94&-0.069&17.893&96.081&14.914&87.152&89.258\\
			Slant down &(i)&438.483&51.864&17.398&5.705&16.767&26.936&-0.069&18.109&95.593&0.304&99.644&100.457\\
			Slant up &(j)&430.807&51.037&23.044&6.79&16.769&26.939&-0.069&20.21&95.261&5.646&99.58&95.023\\
			Star &(k)&247.105&34.388&14.305&13.49&16.769&26.93&-0.063&27.025&86.436&14.366&92.215&80.242\\
			V lines&(l)&344.349&49.928&11.298&3.125&16.77&26.938&-0.069&30.45&89.505&2.735&99.695&98.87\\
			Wide lines &(m)&402.458&44.244&21.591&6.869&16.77&26.938&-0.067&27.44&77.916&0.217&99.284&100.309\\
			X shape & (n) &283.607&32.302&3.603&2.102&16.77&26.93&-0.066&31.107&85.446&4.578&97.838&104.041\\
			
			\hline
		\end{tabular}
	\end{center}
	\raggedright
	\footnotesize{Notes: Table reports summary statistics from the datasaurus dozen \citep{matejka2017same} and a new ``normal'' dataset constructed from two random normal variables with an equal correlation, variance and means as the \cite{matejka2017same} datasets. Persistence norms are denoted by $L_{ab}$ where $a = \lbrace 0,1 \rbrace$ is the dimension and $b = \lbrace 1,2 \rbrace$ is the norm being calculated. All clouds are constructed from two variables, $X_1$ and $X_2$ with identical means, $\mu_1$, $\mu_2$, and standard deviations, $\sigma_1$, $\sigma_2$. For all datasets $\mu_1 = 54.27$, $\mu_2 = 47.84$, $\sigma_1 = 16.77$ and $\sigma_2 = 26.94$. Correlation between the two axis variables, $\rho_{12}$, is also identical in every panel. In all cases $\rho_{12} = -0.064$. The $Dist$ column provides the maximum euclidean distance between two points in the dataset.}
\end{sidewaystable}

Table \ref{tab:norms} shows that the largest values of the dimension 0 $L_1$ norms, $L_{01}$ and $L_{02}$, appear in the Away and Normal datasets, panels (b) and (c), That the Away and Normal datasets have the largest $L_{01}$ and $L_{02}$ norms may be understood from the fact that the points are well spread, with limited clustering. The highest of the maximum distances between points appears for the Normal data set, but there are others such as the X shape in panel (n) which have equally high maximum distance between points. In the case of the X shape, small distances between points on the outside across the X mean that the $L_{01}$ and $L_{02}$ norms stay low. We see the Away dataset in panel (c) having the highest dimension $L_{11}$ and $L_{12}$ norms, but the Normal dataset in panel (b) having low $L_{11}$ and $L_{12}$ norms. The Away dataset has a hole in the centre of the cloud and hence higher dimension $L_{11}$ and $L_{12}$ norms directly linked to that main feature. Likewise, the Bullseye of panel (d) also has a similar $L_{01}$ and $L_{02}$ norms. Panels (b) and (c) also help illustrate that the dimension 0 norms, $L_{01}$ and $L_{02}$, are functions of distance between points, whilst the $L_{11}$ and $L_{12}$ norms are based more on the arrangement of the points.

The third highest $L_{01}$ and $L_{02}$ norms are from the original Dino plot that gives the Datasaurus its name. The Datasaurus does have an arrangement of points around the perimeter of an empty space, this applies to part of the mouth structure and the eye. However, there are features of the Dino which mean that more large shape takes a higher filtration to be born. As these form the $L_{01}$ and $L_{02}$ norms rise. When considering the $L_{11}$ and $L_{12}$ norms of the Dino plot, the lack of closure of the outline for the neck and the connections between the tops of the mouth mean that there are shapes formed by edges only when the points on either side are within twice the filtration radius of each other. The lifetime of these features, and hence the $L_{11}$ and $L_{12}$ norms, are not as long as a result of their later births. The $L_{11}$ persistence norm of the Dino is second only to the Normal dataset, but the $L_{12}$ norm of Dino ranks below the Away and Bullseye datasets, and well below the Circle dataset.

There are then four datasets with $L_{01}$ norms between 400 and 430. These four datasets comprise two of the pairs of rotation plots. Of these Wide lines has a lower $L_{02}$ norm compared to High lines, Slant up or Slant down. Slant down has a lower $L_{11}$ norm, with the other three being very similar. Attribution of the lower norm is to the shape of the data, the maximal distance between points in the dataset being highest for Slant down. On $L_{12}$ norms, we likewise see that Slant down has the lowest norm. Ex ante, the higher maximal distance would suggest higher norms, but specifics of the locations of the points within the \cite{datasauRus2022} package graphs mean the expected relationship does not hold. Inspecting the Slant down and Slant up scatter plots, we see slightly more clustering of points of the Slant up dataset. The difference between Slant up and Slant down is that the former has 5 sets of points in parallel ``slants'', whilst the latter comprises only 4 sets of points with 6 points acting as a very sparse 5th ``slant''. As many of the dimension 1 features form because of connections between the points on neighbouring ``slants'', $L_{11}$ and $L_{12}$ will be lower for the Slant down.

With $L_{01}$ norms between 300 and 400 we find 4 datasets. 2 of the data sets with $L_{01}$ in the range 300 to 400 are rotations, being the H lines and the V lines datasets. We see that the $L_{02}$ norms are very similar to the Slant up, Slant down, Wide lines and High lines datasets. All 6 of the rotation datasets therefore have very similar $L_{02}$ norms to each other. The $L_{11}$ and $L_{12}$ norms for H lines and V lines, $L_{11}$ and $L_{12}$ are much lower than for the other rotation cases. For example the $L_{12}$ norm of High lines is 6.921, but the $L_{12}$ norm of H lines is just 2.558. The difference between H lines and High lines is best understood through looking at the scatterplot in Figure \ref{fig:scatter}. Where H lines has 5 parallel lines of points, High lines has just 2. Where H lines points are tight around the horizontal line, High lines has more spread within the 2 groups of points. Further the distance between the two High lines groups is much larger than any of the gaps between the horizontal sets of points in H lines. Larger gaps and features within the horizontal sets, help to give the High lines dataset higher persistence norms than H lines.

There are 3 data sets with $L_{01}$ norms between 200 and 300, being the Dots, the Star and the X shape. Again these three datasets vary greatly on the other measures. Dots produces an $L_{02}$ norm of 73.71 compared to 34.39 for the Star and 32.02 for the X shape. Dots features 9 small clusters of points, with the potential to form a 2x2 grid of features once the clusters connect. Consequently we see $L_{11}$ at 29.90 and $L_{12}$ being14.63. These high values for Dots compare with 14.31 and 13.49 for the Star and then 3.603 and 2.102 for the X shape. In dimension 1, the central hole in the star is a large feature which adds greatly to the norms. By contrast there are very few shapes to form from the X shape. Because increasing the filtration slightly connects new points on the arm and does not create any dimension 1 features. Visually the Dots, Star and X shape datasets are very different and the persistence norms confirm that difference through the range of persistence norm values observed.

Finally, the Circle has the lowest dimension 0 norms. This may be surprising, but can be rationalised by the fact that all points on the circumference of the circle are close together. Every point can be connected without needing the filtration to be large. In the other cases links between groups of points are needed to reach the state where all data points are connected. However, because the Circle has a large hole in the centre, it has the highest dimension 1 norms. The $L_{12}$ norm of the circle is twice as high as that of the second highest. In this case, the $L_{12}$ norm is reflecting the fact that there is a single long lived dimension 1 feature, rather than there being several smaller dimension 1 features which cease to exist at lower filtrations. Recall $L_2$ norms involve a squared term such that longer-lived features impact the $L_2$ norm more than short-lived features. The large $L_{12}$ norm for the Circle evidences the impact of squaring the lives in dimension 1. The Normal plot added in this paper makes a strong case for the alternative, with a similar $L_{11}$ norm to the Circle, but an $L_{12}$ norm which is around 30\% of that of the Circle.

From the consideration of the 14 datasets, we see that there are different criteria affecting the value of the four norms. We see variation in the norms coming from the appearance of the points within the space. On an intuitive level, the persistence norms are reacting to the joint distribution of the axis variables. In the datasaurus we see that datasets with clusters of points, especially clusters where clusters have identical values on one, or more, of the analysed data

\subsection{Persistence Correlations}

Our discussion of the persistence norms reveals that the ranking of the datasets varies according to which norm is used. We considered four persistence norms alongside the maximum distance between points. As a final exercise let us compare the correlations between the measures. For the correlation we report Pearson and Spearman's rank correlation measures in Table \ref{tab:cora}. Values are based on 14 observations, 1 for each of the 14 datasets.

\begin{table}
 \begin{center}
     \caption{Correlation of Persistence Norms}
     \label{tab:cora}
     \begin{tabular}{cccccccccc}
     \hline

         &$L_{01}$&$L_{02}$&$L_{11}$&$L_{12}$&$MaxDist$\\
         \hline
         $L_{01}$ &1&0.437&0.393&0.077&0.662\\
         $L_{02}$&0.449&1&0.152&-0.020&0.336\\
         $L_{11}$&0.528&0.232&1&0.855&-0.090\\
         $L_{12}$&-0.160&-0.278&0.661&1&-0.341\\
         $MaxDist$&0.730&0.407&0.027&-0.545&1\\
\hline
     \end{tabular}
 \end{center}
 \raggedright
 \footnotesize{Notes: Figures represent correlations between the persistence norms, maximum distances between points, and the skewness and kurtosis measures for the 12 plots of the \cite{datasauRus2022} package, the original daatsaurus plot and the normal distribution plot. The $MaxDist$ column provides the maximum euclidean distance between two points in the dataset. All clouds are constructed from two variables, $X_1$ and $X_2$ with identical means, $\mu_1$, $\mu_2$, and standard deviations, $\sigma_1$, $\sigma_2$. For all datasets $\mu_1 = 54.27$, $\mu_2 = 47.84$, $\sigma_1 = 16.77$ and $\sigma_2 = 26.94$. Correlation between the two axis variables, $\rho_{12}$, is also identical in every panel. In all cases $\rho_{12} = -0.064$. The Normal dataset is constructed with the code accompanying this paper. Full details of the datasets can be seen in Section 3 of this note.}
\end{table}

From Table \ref{tab:cora} we see a formalisation of the discussion of the individual datasets. Dimension 0 $L_1$ norms are strongly correlated with the maximum distance between two points in the dataset; the Pearson correlation is 0.730. There is then some correlation with the dimension 0 $L_2$ norm, but we see that the Pearson correlation betwen $L_{01}$ and $L_{02}$ is just 0.449. The Pearson correlation between $L_{02}$ and the maximum distance between points is also notable at 0.407. The highest correlation observed in the data is between the dimension 1 $L_1$ and $L_2$ norms. Spearman's rank correlation coefficient for $L_{11}$ and $L_{12}$ is 0.855, while the Pearson correlation is also high at 0.661. $L_{11}$ is not correlated with the maximum distance between any pair of points, but $L_{12}$ does show a negative correlation with $MaxDist$. The primary message from Table \ref{tab:cora} is that there is limited correlation between the TDA measures and the maximum distance between points.  

\section{Extensions}

The datasaurus dozen are well defined, but the inference drawn upon their persistence norms may be further understood by considering transformations of the data and by looking at the univariate distributions of each data set. We present results from scaling of the point clouds, translation and when the data sets are stretched on a single axis. The univariate distributions are presented to remind that whilst the first and second moments of the datasaurus dozen are identical, the third and fourth moments are not. 

\subsection{Transformations of the Data}

All TDA invariants are invariants of affine transformations. In the case of the plots used in this note TDA is invariant to isometries. To demonstrate, three transformations are considered. Scaling the data sets by multiplying $X_1$ and $X_2$ by a constant $\alpha$ produces higher norms, consistent with the fact that scaling alters the distance matrix but does not alter the relative locations of points. A second transformation is achieved by shifting the data points by $\delta$. Translation has no effect on the persistence norms as the distance matrix is unaffected when all points shift by the same amount. Finally, we consider the multiplication of $X_1$ by a constant $\beta$, but keeping $X_2$ unchanged. This third shift changes the distance matrix and the relative positions of points within the cloud. Changes in the norms from scaling one variable are therefore different in each data set.

\begin{figure}
    \begin{center}
        \caption{Dataset Transformations}
        \label{fig:trans1}
        \begin{tabular}{c c c}
             \includegraphics[width=5cm]{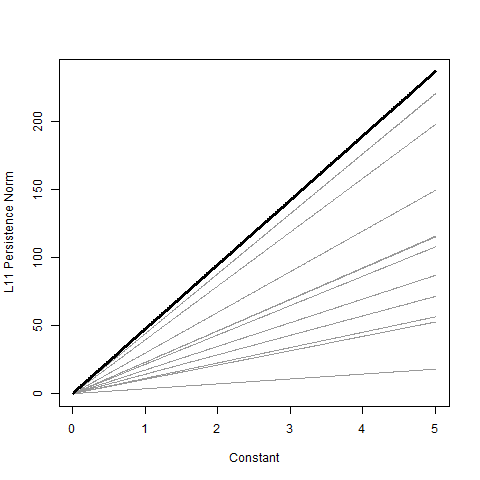}&
             \includegraphics[width=5cm]{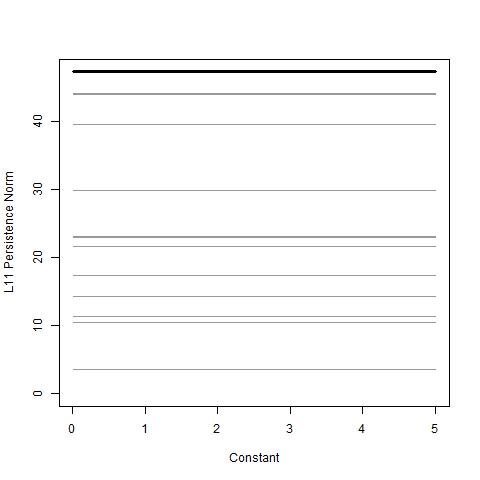}&
             \includegraphics[width=5cm]{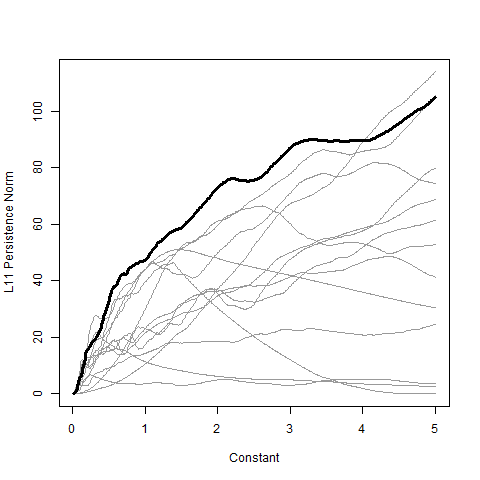}\\
             (a) Scaling $L_{11}$ & (b) Translation $L_{11}$ & (c) Expansion $X_1$ $L_{11}$\\
             \includegraphics[width=5cm]{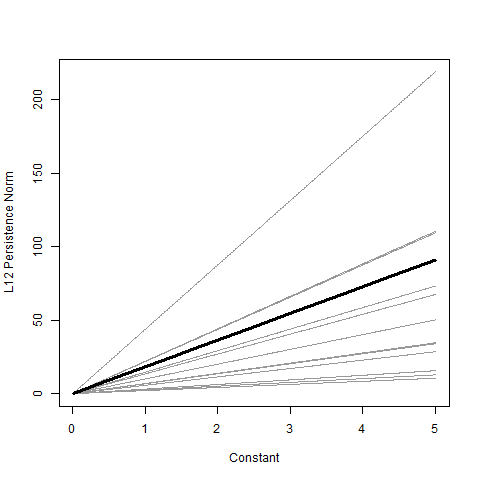}&
             \includegraphics[width=5cm]{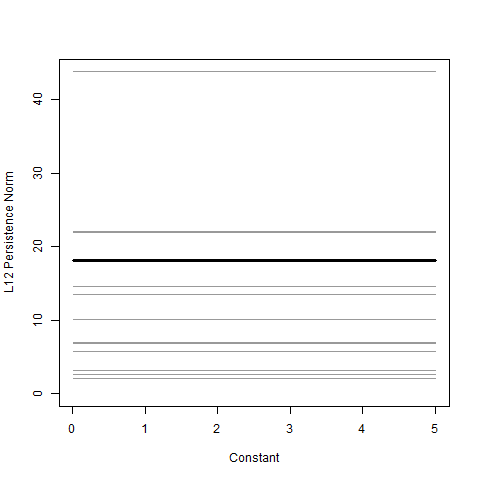}&
             \includegraphics[width=5cm]{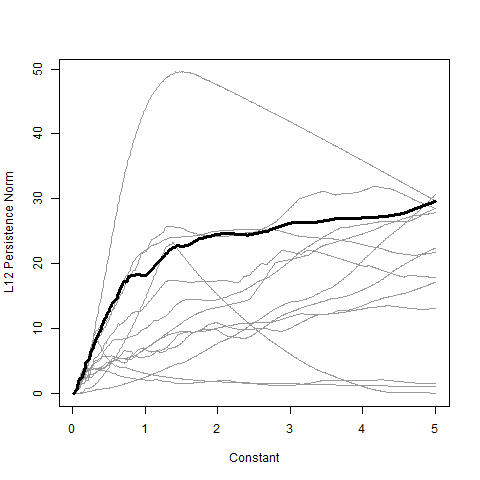}\\
             (d) Scaling $L_{12}$ & (e) Translation $L_{12}$ & (f) Expansion $X_1$ $L_{12}$\\
        \end{tabular}
    \end{center}
    \raggedright
    \footnotesize{Notes: Figures plot impact of three transformations to the points in the original datasaurus and the datasaurus dozen of \cite{matejka2017same}. A 14th dataset based on normal distributions is also created for this paper and included in the analysis. Panels (a) and (d) show the impact of multiplying all points by a constant to expand the cloud. Panels (b) and (e) show the effect of translating the cloud along the $X_1$ axis by adding a constant. Panels (c) and (f) show the effect of multiplying the $X_1$ co-ordinates by a constant but not changing the $X_2$ co-ordinates. In all cases there is one grey line for each dataset. The solid black line represents the original datasaurus.}
\end{figure}

Figure \ref{fig:trans1} has 6 panels. Panels (a) and (d) concern the first transformation, a scaling of the dataset by a constant $\alpha$. On the horizontal axis we plot $\alpha$ and on the vertical axis we depict the persistence norms. The impact in every case is linear, with the slope of the line depending on the norm observed when $\alpha=1$ in the previous section.  Panels (b) and (e) show the effect of translation by $\delta$ to be 0; persitence norms are constant across all $\delta$. Panels (c) and (f) represent the greatest interest since we see seemingly unordered behaviour when the $X_1$ co-ordinates of all points are scaled by $\beta$. In the case of the dinosaur, shown as a thicker black line, the norms increase with $\beta$, but other datasets show hump shapes and have decreasing norms for higher $\beta$. Lines in panels (c) and (f) cross many times meaning that the ranking of the norms of the 14 data sets are constantly changing with $\beta$. 


\subsection{Distributions of the Datasaurus}

\begin{figure}
	\begin{center}
		\caption{Axis Distribution Plots}
		\label{fig:dists}
		\begin{tabular}{c c}
			\includegraphics[width=5cm]{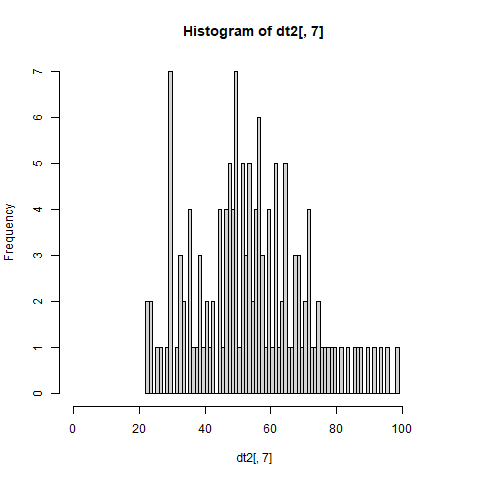}&

			\includegraphics[width=5cm]{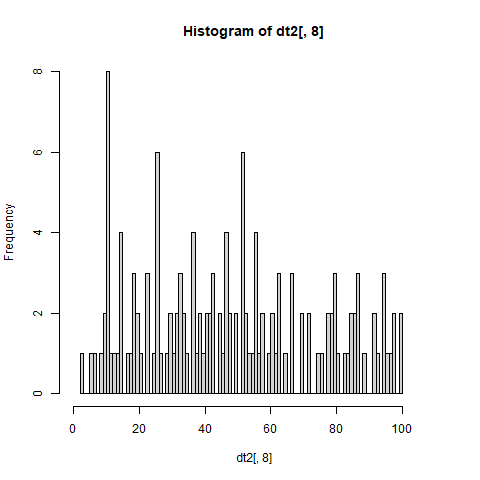}\\
			(a) Dinosaur $X_1$ & (b) Dinosaur $X_2$ \\
   			\includegraphics[width=6cm]{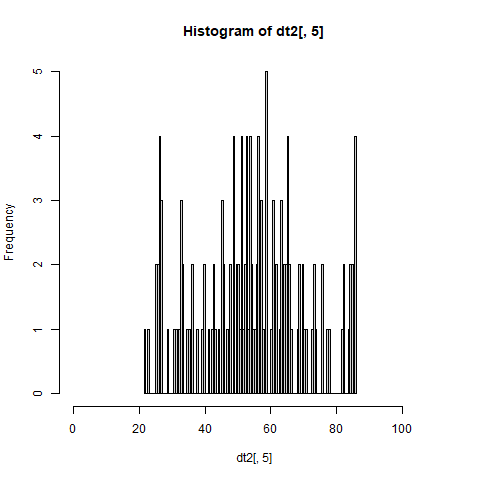}&
			\includegraphics[width=6cm]{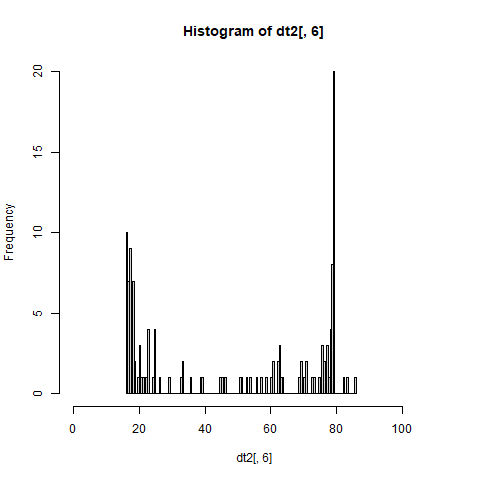}\\
			(c) Circle $X_1$ & (d) Circle $X_2$ \\
   			\includegraphics[width=6cm]{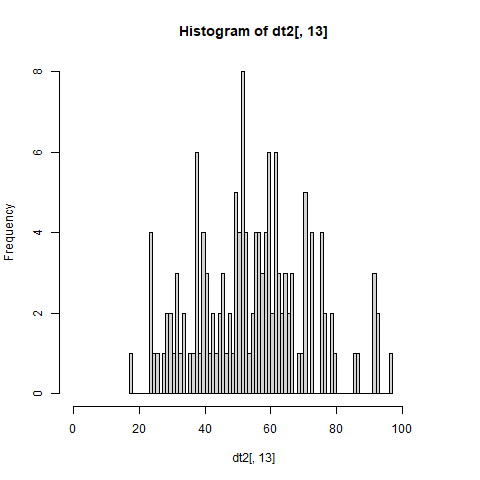}&
			\includegraphics[width=6cm]{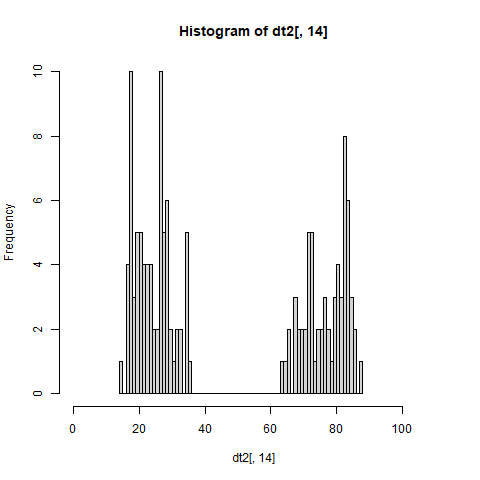}\\
			(e) High Lines $X_1$ & (f) High Lines $X_2$ \\
		\end{tabular}
	\end{center}
\raggedright
\footnotesize{Notes: Figures plot densities of $X_1$ and $X_2$ variables in selected datasets from the datasaurus dozen of \cite{matejka2017same}. Panels (a) and (b) refer to the original Dinosaur, (c) and (d) are from the Circle, and (e) and (f) are based upon the High Lines dataset.}
\end{figure}

Figure \ref{fig:dists} plots histograms of the densities of $X_1$ and $X_2$ for three of the datasets considered in this note. The left column contains plots for $X_1$ and the right column shows $X_2$. Immediately we see the difference between the datasets. Dinosaur, panels (a) and (b) has a spread of values for both variables, albeit that $X_1$ displays a humped distribution around the mean. $X_2$ shows greater spread generating a more sawtooth looking plot. Panels (c) and (d) give indication as to the source of the high $L_2$ norm for the circle. The twin peak of the distribution of $X_2$ contrasts with the more normal shaped $X_1$. The peaks on $X_2$ correspond to where the circle is near horizontal on the plot. Closer inspection shows that the number of points in these horizontal sections are much higher than at the end where the circle becomes more vertical. Finally, panels (e) and (f) show the two peaks for the wide lines datasets. The higher norms in the High Lines case formed when connections appeared between the two ``lines'' on the plot. On the distribution we see each ``line'' as a denser part of the $X_1$ variable. Like the Circle there are clear peaks, but relative to the Circle those peaks are closer on the $X_2$ axis. 

Comparison of the Circle and High Lines case suggests that the existence of twin peaks in the distribution on one axis is a signal of higher norms. However, consider the case of a ``H'' shaped plot. On the horizontal dimension there are two peaks, but on the vertical axis there is only a single peak, corresponding to the horizontal bar. The letter ``H'' has no holes and therefore $L_2$ norms will be close to zero\footnote{If the data is perfectly shaped as a ``H'' with regularly spaced points there will be a 0 $L_2$ norm, but small variations open the possibility of very small non-zero $L_2$ norms appearing.}. Consequently the twin peaks in the distribution of one variable are suggestive of higher norms, but do not ensure higher norms. Norms remain representative of the joint distribution of the data and do not correspond to any one metric of the underlying point cloud.

\begin{table}
	\begin{center}
            \caption{Skewness and Kurtosis}
            \label{tab:sk}
		\begin{tabular}{l c c c c}
		\hline
		Dataset & $Sk_1$&$Sk_2$&$Kt_1$&$Kt_2$\\
		\hline
		Dino&0.28&0.25&2.75&1.96\\
		Normal&0.06&0.06&2.67&2.03\\
		Away&-0.02&-0.09&1.78&1.74\\
		Bullseye&0.17&0.01&2.77&1.57\\
		Circle&0.03&0.00&2.36&1.21\\
		Dots&-0.12&0.15&2.16&1.61\\
		H Lines&0.24&0.15&2.68&1.87\\
		High Lines&0.19&0.20&2.76&1.21\\
		Slant Down&0.25&0.19&2.78&2.08\\
		Slant Up&0.23&0.26&2.65&1.93\\
		Star&0.08&0.24&1.84&1.66\\
		V Lines&0.26&0.23&2.45&1.98\\
		Wide Lines&-0.32&0.25&1.30&1.94\\
		X Shape&0.48&0.23&1.65&1.58\\
		\hline
	\end{tabular}
\end{center}
\raggedright
\footnotesize{Notes: Table reports summary statistics from the datasaurus dozen \citep{matejka2017same} and a new ``Normal'' dataset constructed from two random normal variables with an equal correlation, variance and means as the \cite{matejka2017same} datasets. $Sk_1$ and $Sk_2$ provide the skewness of $X_1$ and $X_2$ respectively. $Kt_1$ and $Kt_2$ provide the kurtosis of $X_1$ and $X_2$ respectively. All clouds are constructed from two variables, $X_1$ and $X_2$ with identical means, $\mu_1$, $\mu_2$, and standard deviations, $\sigma_1$, $\sigma_2$. For all datasets $\mu_1 = 54.27$, $\mu_2 = 47.84$, $\sigma_1 = 16.77$ and $\sigma_2 = 26.94$. Correlation between the two axis variables, $\rho_{12}$, is also identical in every panel. In all cases $\rho_{12} = -0.064$. The Normal dataset is constructed with the code accompanying this paper. Full details of the datasets can be seen in Section 3 of this note.}
\end{table}

More information about the distributions is obtained by considering the third and fourth moments. Table \ref{tab:sk} provides the skewness and kurtosis for each variable within the 14 datasets. Skewness, $Sk_1$ and $Sk_2$, measures the extent to which the density of observations is higher, or lower, than the mean. A normal distribution has a skewness of 0. The 14 datasets have skewness ranging from -0.32 to 0.48, many of the axes have a skew of 0.20. The Circle has the closest skewness to 0, the Normal dataset having $Sk_1$ and $Sk_2$ of 0.06 to 2 decimal places. Kurtosis also varies across the datasets, no two datasets having identical values. A close match is found between Slant Up and Dino, $Sk_1$ being 0.28 for Dino and 0.24 for Slant Up, the $Sk_2$ values are 0.25 and 0.26 for Dino and Slant Up respectively, the $Kt_1$ are 2.75 and 2.65 respectively, and the $Kt_2$ are 1.96 and 1.93. However, we know the persistence norms for the Dino are $L_{01} = 485.7$, $L_{02}= 43.62$, $L_{11} = 47.33$ and $L_{12}=18.16$. For the Slant Up the corresponding persistence norms are $L_{01} = 430.8$, $L_{02} = 51.04$, $L_{11} = 23.04$ and $L_{12}=6.79$. The difference in the dimension 1 persistence norms is particularly pronounced. Whilst the third and fourth moments do reveal differences between the datasets, there are still further elements of the persistence norms which create difference even when the first four moments of two datasets are similar. 

\begin{table}
	\begin{center}
		\caption{Correlation of Persistence Norms}
		\label{tab:corb}
		\begin{tabular}{cccccccccc}
			\hline

			&$L_{01}$&$L_{02}$&$L_{11}$&$L_{12}$&$Dist$& $Sk_1$&$Sk_2$&$Kt_1$&$Kt_2$\\
			\hline
			$L_{01}$&1&0.437&0.393&0.077&0.662&-0.029&0.147&0.341&0.587\\
			$L_{02}$&0.449&1&0.152&-0.020&0.336&-0.319&-0.305&0.209&0.367\\
			$L_{11}$&0.528&0.232&1&0.855&-0.090&-0.459&-0.279&0.226&-0.059\\
			$L_{12}$&-0.160&-0.278&0.661&1&-0.341&-0.534&-0.486&0.002&-0.349\\
			$MaxDist$&0.730&0.407&0.027&-0.545&1&0.068&0.222&-0.222&0.657\\
			$Sk_1$&-0.222&-0.369&-0.491&-0.291&-0.071&1&0.248&0.385&0.121\\
			$Sk_2$&-0.043&-0.057&-0.427&-0.676&0.261&0.162&1&-0.147&0.415\\
			$Kt_1$&0.220&0.218&0.258&0.038&-0.193&0.381&0.014&1&0.156\\
			$Kt_2$&0.531&0.359&-0.001&-0.489&0.663&-0.084&0.406&0.061&1\\
			\hline
		\end{tabular}
	\end{center}
	\raggedright
	\footnotesize{Notes: Figures represent correlations between the persistence norms, maximum distances between points, and the skewness and kurtosis measures for the 12 plots of the \cite{datasauRus2022} package, the original daatsaurus plot and the normal distribution plot.he $Dist$ column provides the maximum euclidean distance between two points in the dataset. $Sk_1$ and $Sk_2$ provide the skewness of $X_1$ and $X_2$ respectively. $Kt_1$ and $Kt_2$ provide the kurtosis of $X_1$ and $X_2$ respectively. All clouds are constructed from two variables, $X_1$ and $X_2$ with identical means, $\mu_1$, $\mu_2$, and standard deviations, $\sigma_1$, $\sigma_2$. For all datasets $\mu_1 = 54.27$, $\mu_2 = 47.84$, $\sigma_1 = 16.77$ and $\sigma_2 = 26.94$. Correlation between the two axis variables, $\rho_{12}$, is also identical in every panel. In all cases $\rho_{12} = -0.064$. The Normal dataset is constructed with the code accompanying this paper. Full details of the datasets can be seen in Section 3 of this note.}
\end{table}

Adding skewness and kurtosis to the correlation matrix, Table \ref{tab:corb} shows that there is a stronger correlation betwen skewness, $Sk_1$ and $Sk_2$, and the dimension 1 persistence norms $L_{11}$ and $L_{12}$. Kurtosis, $Kt_1$ and $Kt_2$ have stronger correlations with the dimension 0 persistence norms $L_{01}$ and $L_{02}$. There is a strong negative correlation between $Kt_2$ and $L_{12}$. Ex-ante there is no reason for positive, or negative skew to create different persistence norms. The correlation observed here is a function of the specific datasets and reminds that we are only considering a sample of 14 observations. Kurtosis captures the peakedness of the distribution, lower kurtosis means more spread of observations across the range of the $X$ variable. in this case lower kurtosis would be expected to produce higher norms; consistent with the negative correlations in Table \ref{tab:corb}. Again differences between $X_1$ and $X_2$ are products of the specific datasets rather than being of statistical interest.

\section{Discussion}
\label{sec:discuss}

The purpose of this note is a reminder that whilst data may indeed have the same \PD{first and} second order summary statistics, differentials in the joint distribution of the data points can, in some cases, be picked up by TDA-based statistics. TDA is playing the role of visualising the data, identifying patterns and shapes without actually showing the data points. Mapper algorithms, such as TDA BallMapper~\citep{dlotko2019ball} can be used to plot multi-dimensional datasets and allow the extension of the results of \cite{matejka2017same} into multiple dimensions. The application of TDA BallMapper is left to further work. Our aim in this note is solely to demonstrate that different persistence norms may emerge from datasets with identical correlations and standard deviations. With the further objective to say more about the motivations for the differences in persistence norms.

Although it is interesting to think about datasets with patterns, the majority of applications that we encounter in applying statistics do have distributions which are close to normal. A pattern like the Dots example is unlikely to apply in practice. In practice, the consistent message from the study of the datasaurus is that the formation of groups of points within a larger dataset, or the creation of large voids within the space over which data is distributed, will be associated with higher norms. Such patterns are picked up most by the $L_{12}$ persistence norm.

Implications from this note to the literature on persistence norms and the features of time series emerge. Firstly, the desire to explain persistence norms through only the standard deviation of the values of the time series, and from the correlations between time series, is laudable but requires the assumption that the distributions of the observations are consistent over time. For example if considering the returns on two stocks, it must be assumed that both stocks have close to normal distributions of returns. Spread of returns, and correlations between the stocks may change, but the fact that the distribution is approximately normal should not. When the underlying distribution changes, for example becomes skewed, then the persistence norms will pick up the difference. Where the assumption of identical distributions is reasonable, movements in the persistence norms not associated with changes in correlation or standard deviation, may therefore indicate the appearance of clustering within the data\footnote{Recent related work by \cite{dlotko2022topology} presents a TDA-based test for two samples being from the same joint distribution. During the process of perturbation which produces the example point clouds used in this note, \cite{matejka2017same} does not test that the points are from the same underlying joint distribution.}. The consequences of clustering of points within the cloud then depend on the specific dataset to which TDA is being applied.

There remains an open research challenge to identify the consistent determinants of persistence norms. The answer will be specific to the data being analysed and the underlying distributions thereof. TDA unlocks information from the joint distribution of multiple continuous variables. A long conversation between the advancement of TDA, and the applications in specific disciplines, remains to be had.

\bibliography{datasaurus}
\bibliographystyle{apalike}

\end{document}